\documentclass{ws-p10x7}

\begin{document}
\newcommand{\gsim}{\raisebox{-0.5mm}{$\stackrel{>}{\scriptstyle{\sim}}$}} 
\newcommand{\lsim}{\raisebox{-0.5mm}{$\stackrel{<}{\scriptstyle{\sim}}$}}   

\title{Jet Production in DIS at HERA}

\author{J\"org Gayler}

\address{DESY, Hamburg, Germany
 \\E-mail: joerg.gayler@desy.de}

\twocolumn[\maketitle\abstract{
 Data on jet production in deep inelastic $e^+p$ scattering are presented.
 The results are compared with pQCD calculations. At low $Q^2$ no
 consistent description of the data over all the phase space is
 available yet. At high $Q^2$ ($\gsim 150$ GeV$^2$) the data are well
 described by pQCD in NLO.
                                         }]

\section{Introduction}

 Inclusive deep inelastic lepton nucleon scattering, where only the 
 scattered lepton is detected, played an important role in
 establishing QCD and continues to provide a well
 defined testing ground of perturbative QCD (pQCD).
 The aim of measurements
 of final state jets is to relate them to final state quarks and gluons
 and thereby to gain additional insight in the dynamics of lepton nucleon
 scattering.

 The data presented in this talk\footnote{ICHEP2000, Osaka, 2000}
 were recorded in the years 1995 to 1997
  at HERA with the H1 and ZEUS detectors
 where positrons of 27.5 GeV collided with protons of 820 GeV. 

\subsection{Kinematics}\label{subsec:kin}

\begin{figure}[htb]
\begin{center}
  \epsfig{file=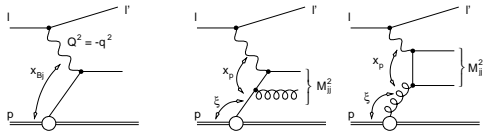,bbllx=80pt,bblly=644pt,bburx=525pt,bbury=774pt,width=6.6cm,clip=}
\end{center}
\caption{Processes in DIS: Born process; QCD-Compton process; and boson-gluon
  fusion (left to right).}
\label{fig:kin1}
\end{figure}

 The basic Feynman diagrams describing jet production in deep inelastic 
 scattering are shown in Fig.~\ref{fig:kin1}.

 Standard kinematic quantities
 \footnote{Polar angles $\theta$ are measured with respect to the incident
  proton direction, the pseudo rapidity is given by
  $\eta = -\ln(\tan\theta/2)$.} 
 are
 $ Q^2 = -q^2 = -(l-l')^2$,
  the virtuality of the boson exchange
 and the Bjorken variable
 $x_{Bj} = Q^2/2 pq$.
 The momentum fraction entering the hard process of jet production with
 a jet-jet mass $M_{jj}$ (see Fig.~\ref{fig:kin1}) is given by 
 $\xi = x_{Bj} (1 + M^2_{jj}/Q^2)$ of which the fraction 
      $x_p = x_{Bj}/\xi$ 
 interacts with the exchanged boson.

 In most cases 
 the data are
 analysed in the Breit frame defined by the condition         
   $ 2 x_{Bj}  \vec p +  \vec q = 0$.
 Quark parton model like events (Fig.~\ref{fig:kin1}, left) exhibit no $p_t$
 in this frame apart from effects of fragmentation and decays.
 Jet finding is performed mostly using the 
 inclusive $k_t$ algorithm~\cite{incl}.

\subsection{Multi-Jet Production in pQCD}\label{subsec:pQCD}

  Calculations at the parton level are available up to order $\alpha_s^2$,
  i.e. to next to leading
  order (NLO)
 (Fig.~\ref{fig:kin1} shows diagrams up to leading order (LO)).
  They can be compared with data after corrections
  for hadronisation are applied. DISENT~\cite{dis} and DISASTER++~\cite{disas}
  have been shown~\cite{Dup}
  to agree in the kinematic range of interest here.
  MEPJET~\cite{mep} is the only program implementing also
  charged current reactions and
  JetVip~\cite{jetv} allows resolved photon processes to be included. 

  A common ambiguity in these fixed order calculations is the choice of the
  renormalization scale $\mu_R^2$.
  Typical quantities characterizing the process are
  $Q^2$ and $E_t^2$ and the agreement with the data for these hard scales
  and the sensitivity to scale variations is studied.

  Forward jets, i.e. jets close to the proton remnant, are of special
  interest, because they are expected~\cite{muel} to be sensitive probes
  of the evolution of parton densities.
  In particular in $~\alpha_s \log(1/x)$ resummation
  (BFKL approach) one expects
  jets with larger $p_t$ (``$k_t$'') close to the proton remnant than
  in the standard $\alpha_s \log(Q^2)$ resummation (DGLAP approach),
  due to the strong $k_t$ ordering in the latter case. 

\section{$\phi$ Asymmetries}

 A measurement of the $\phi$ distribution 
 of charged particle tracks has been presented by the ZEUS~\cite{zphi}
 collaboration,
for different transverse momentum cuts.
 Here $\phi$ is the azimuthal angle
 of the hadron production plane with respect
 to the positron scattering plane in the hadronic centre of mass system.
 Finite terms $B < 0$ and $C > 0$ were
 measured in the angular distribution
$d\sigma/d\phi = A + B \cos(\phi) + C \cos(2\phi)$ as expected in
 QCD-based calculations.

\begin{figure}[htb]
\begin{center}
\epsfig{file=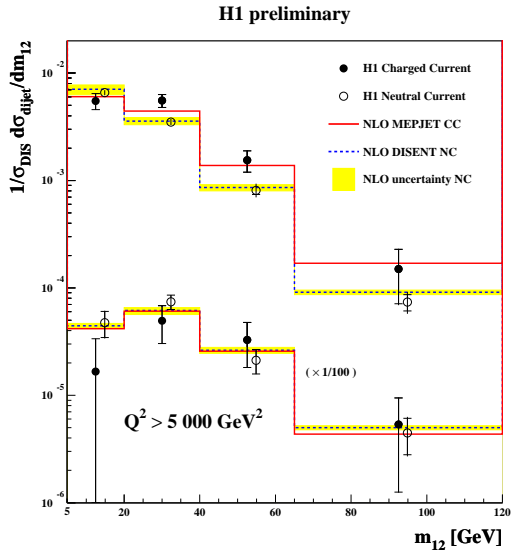,width=6.8cm,clip=}
\end{center}
\caption{Distribution of jet-jet mass $m_{12}$
 (not including the proton remnant jet)
  for CC and NC 
   interactions. 
 $p_t^{lepton} > 25$ GeV,
    $Q^2 > 640$ GeV$^2$ and $Q^2 > 5000$ GeV$^2$ (the latter scaled by 1/100).}
\label{fig:CC}
\end{figure}
\section{Jets in CC Interactions} 

 Jet distributions
  in charged current (CC) interactions
 at high $Q^2$
  are consistent with 
 pQCD expectations
 (see Fig.~\ref{fig:CC}).
The differences to neutral current (NC) jets
are mainly due to the different boson propagators~\cite{h1cc}.  
 
\section{Jets at Low and High $Q^2$} 

\begin{figure}[htb]
\begin{center}
\epsfig{file=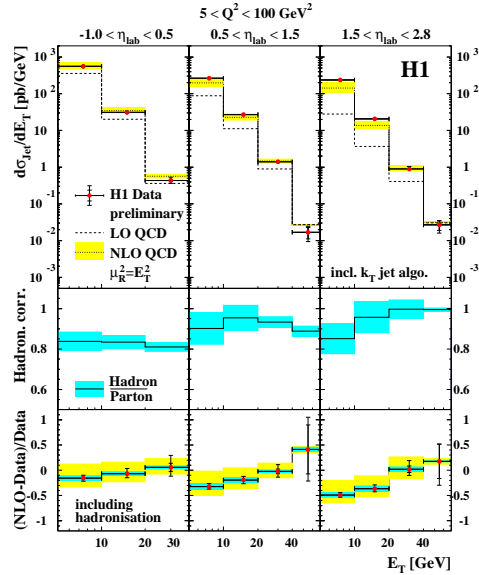,width=6.1cm,clip=}
\end{center}
\caption{$d\sigma_{Jet}/dE_T$ compared to LO (dashed) and NLO (dotted) 
  pQCD (DISENT) predictions using $\mu^2_R=E_T^2$. The shaded band
  shows the sensitivity to scale variations by a factor 4. Also shown are
  the hadronization corrections and the relative deviations after their application.}
\label{fig:f3}
\end{figure}

\begin{figure}[htb]
\begin{center}
\epsfig{file=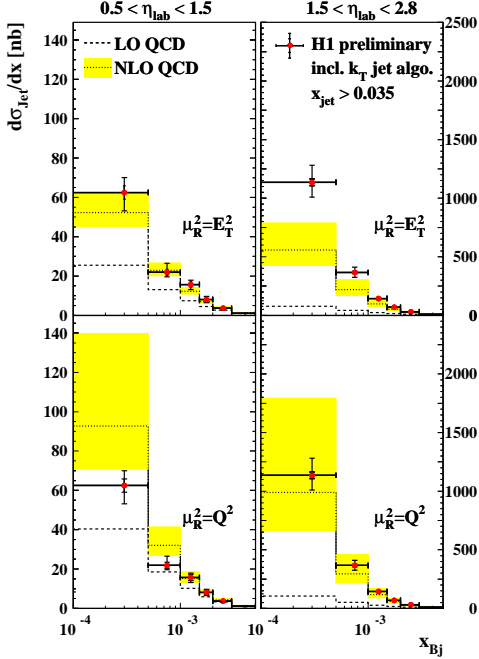,width=6.3cm,clip=}
\end{center}
\caption{$d\sigma_{Jet}/dx$ for $Q^2 > 5 $GeV$^2$ in two regions of $\eta_{lab}$.
  NLO calculation in upper plots with $\mu^2_R=E_T^2$, in lower with $\mu^2_R=Q^2$.}
\label{fig:xbj}  
\end{figure}

 The $E_T$ distribution in the Breit frame of single-inclusive jets is shown 
 in Fig.~\ref{fig:f3} in different regions of $\eta_{lab}$.
 The discrepancies visible in the forward region,
 where the NLO corrections are huge, originate predominantly from small $Q^2$.
 The $x_{Bj}$ dependence in the forward and central region (Fig.~\ref{fig:xbj})
 cannot be described
 with the scale $\mu^2_R=E_T^2$. A consistent description is possible 
 with $\mu^2_R=Q^2$, but the susceptibility to scale variations is vastly
 increased 
 (shaded band in Fig.~\ref{fig:xbj})~\cite{h1low}.

\begin{figure}[htb]
\begin{center}
\epsfig{file=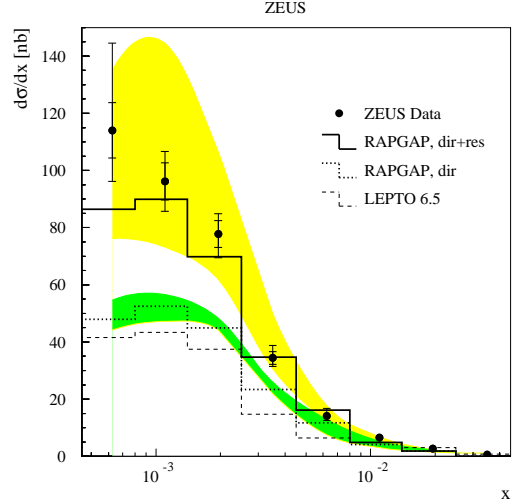,width=6.7cm,clip=}
\end{center}
\caption{$d\sigma_{Jet}/dx$ for 
   p$_z^{Breit} > 0$ , $Q^2 > 10$ GeV$^2$ , $0.5 < E^2_T/Q^2 < 2$. Comparisons
   with models of direct
   (LEPTO, RAPGAP,dir) and direct and resolved
   photon interactions (RAPGAP, dir+res).}
\label{fig:zxbj}
\end{figure}

 Such forward cross sections can be described by the NLO program JetVip and 
 by DGLAP-based QCD Monte Carlo models
 if the hadronic 
 structure of the interacting virtual photon is resolved
 (RAPGAP~\cite{rapgap}, dir+res   
 in Fig.~\ref{fig:zxbj}),
 whereas inclusion of direct photon interactions only
 (RAPGAP, dir and LEPTO~\cite{lepto})
 is insufficient~\cite{z896}.
 In the case of resolved photons, the strong $k_t$ ordering
 is effectively lost,
 leading to larger jet $E_T$ close to the proton remnant.
 However, there are ambiguities in JetVip in the treatment of
 parton 
 masses and no general solution has been found which is
 consistent with the H1 data in a large range of
 rapidities $\eta_{lab}$~\cite{h1low}.

 For detailed discussions of di-jet production at low $Q^2$ see the
 contributions~\cite{h1dilow,zdilow}.
 
\begin{figure}[htb]
\begin{center}
\epsfig{file=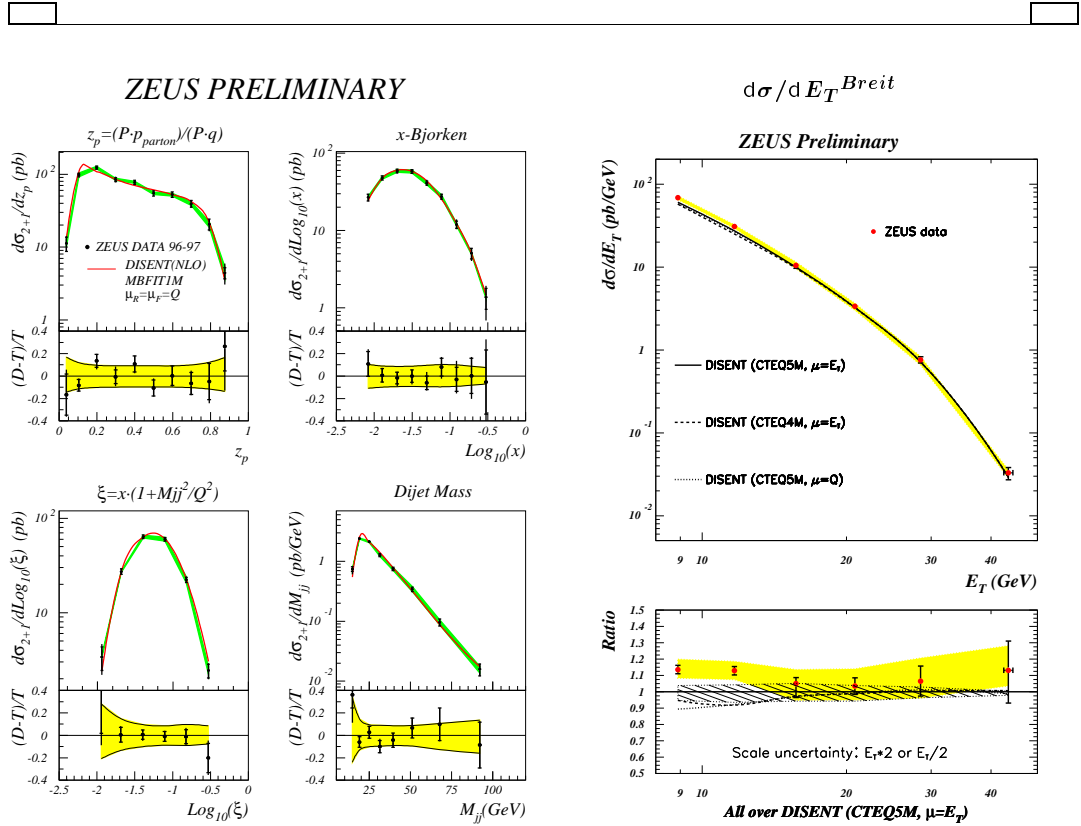,width=6.9cm,clip=}
\end{center}
\caption{Di-jet distributions for
$470 < $Q$^2 < 20000$ GeV$^2$, $-1 < \eta^{lab}_{jet} < 2$,
 E$_T^{jet1} > 8$ GeV and E$_T^{jet2} > 5$ GeV.
 The upper band shows the experimental energy scale uncertainty, the 
 lower the uncertainty of the NLO calculation.}
\label{fig:zdi}
\end{figure}

\begin{figure}[htb]
\begin{center}
\epsfig{file=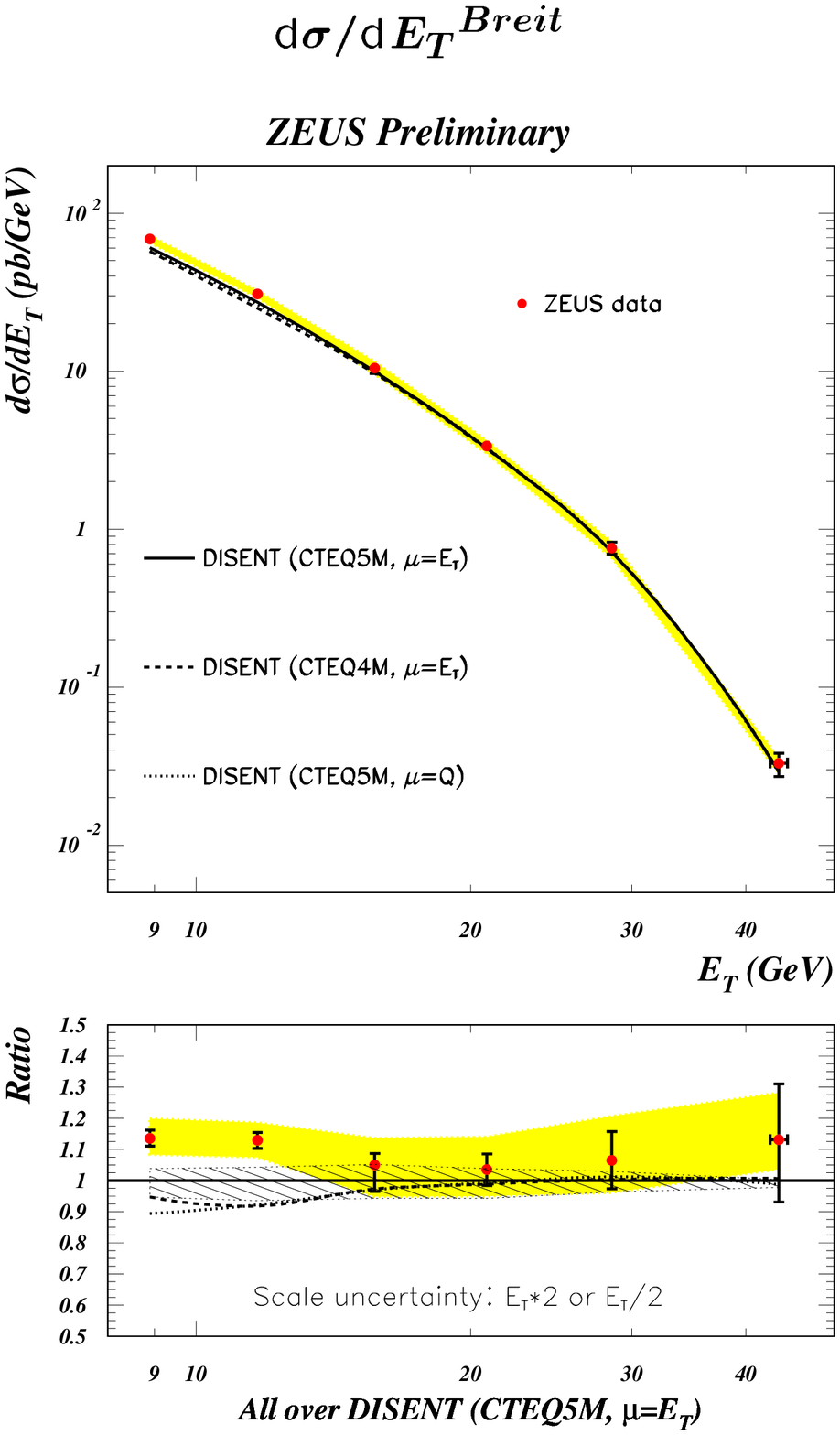,width=5.9cm}
\end{center}
\caption{Inclusive jets, Q$^2 > 125$ GeV$^2$, 
  \newline $-2 < \eta^{Breit}_{jet} < 1.8$}
\label{fig:zincl}
\end{figure}

  At high $Q^2$ there are precise high statistics data available from 
  H1 and ZEUS which agree with NLO calculations on the 10\% level
  in detailed comparisons  
(see Fig.~\ref{fig:zdi})~\cite{zdihi}. For inclusive jets ZEUS
  reports~\cite{zinhi} at $Q^2 < 250$ GeV$^2$ some disagreement on the 15\%
  level for $E_T^2$ and $Q^2$ scales (see Fig.~\ref{fig:zincl}),
 but otherwise the agreement of data and NLO calculations (DISENT)
  is very good~\cite{h1fit}.  

\section{Conclusion}

   The description of the available jet data is considerably improved
   in going from LO to NLO
   ($\sim \alpha_s^2$) pQCD. However some definite
   discrepancies remain to be resolved. 
   They are more pronounced choosing 
   $E_T^2$ as renormalization scale than for $Q^2$. In the latter case 
   the effects of scale variations are large. Forward jets
   are better described if the hadronic structure
   of the virtual photon is taken into account.  
   At high $Q^2$ ($\gsim 150$ GeV$^2$) the data are well described by NLO pQCD,
   the NLO corrections are moderate and hadronization corrections
   $\lsim 10$\%.
   These data are well suited for quantitative QCD analyses~\cite{tassi}.

\section*{Acknowledgments}
 I am grateful to T. Sch\"orner and M. Wobisch
 for discussions and to E.~Elsen and B.~Foster 
 for comments on the manuscript.

\end{document}